# SOBRE LA HABITABILIDAD DE EXOPLANETAS ORBITANDO A PRÓXIMA DEL CENTAURO

ON THE HABITABILITY OF EXOPLANETS ORBITING PROXIMA CENTAURI

M. López-Águila, R. Cárdenas-Ortiz y L. Rodríguez-López

Laboratorio de Ciencia Planetaria. Dpto. Física, Universidad Central "Marta Abreu" de Las Villas, Santa Clara. Cuba

En este trabajo se aplica un modelo matemático de fotosíntesis para estimar la habitabilidad de un planeta hipotético orbitando Próxima del Centauro dentro de la zona de habitabilidad. Los resultados sugieren viabilidad apreciable para la productividad biológica, si los organismos presentes han evolucionado hasta desarrollar la capacidad de utilizar luz infrarroja para la fotosíntesis.

We apply a mathematical model for photosynthesis to quantitatively assess the habitability of a hypothetical planet orbiting Proxima Centauri, inside the so called habitability zone. Results suggest significant viability for primary biological productivity, provided living organisms have evolved to reach the ability of using infrared light for photosynthesis.

PACS: Astrobiology, *91.62.Fc and 96.55.+z. Photosynthesis in oceanography, 92.20.Cm and *92.20.ch

## I. INTRODUCCIÓN

Alrededor del 75% de las estrellas de la Vía Láctea son enanas rojas [1]. Como éstas son pequeñas y frías, tienen la zona de habitabilidad en un rango pequeño y es posible que los planetas que se encuentren orbitando en esta zona estén en anclaje de marea [2,3], lo cual implica potenciales complicaciones climáticas debidas al drenaje de energía térmica del hemisferio iluminado al oscuro. A esta último se suman las frecuentes emisiones de rayos X [4]. Todo lo anterior sugiere que la probabilidad de albergar vida en planetas alrededor de estas estrellas es baja. Sin embargo, dada su abundancia, se continúan estudios en esta dirección [1].

Próxima del Centauro es la estrella más cercana al Sol, y por esto es estudiada por varios observatorios [5]. Es una estrella enana roja y el 85% de la radiación que emite es en la parte del espectro que corresponde al infrarrojo [6].

En este trabajo se calculan las tasas de fotosíntesis del fitoplancton en un océano en calma y en un océano con corrientes verticales circulares (circulación de Langmuir) de un planeta hipotético que orbita en la zona habitable de Próxima del Centauro.

## II. MATERIALES Y MÉTODOS

*2.1. Modelo de emisión estelar*   El espectro de emisión de una estrella se puede aproximar al del cuerpo negro, por ello se hace uso de la ley de Planck para obtener las irradiancias espectrales $E_{star}(\lambda)$ emitidas por la estrella:

$$E_{star}(\lambda) = \frac{2\pi h c^2}{\lambda^5} \frac{1}{e^{\frac{hc}{k_B T \lambda}} - 1}, \qquad (2.1)$$

donde $h$ es la constante de Planck, $c$ es la velocidad de la luz en el vacío, $\lambda$ es la longitud de onda de la radiación, $k_B$ es la constante de Boltzmann y $T$ es la temperatura efectiva de la estrella (3042±117 K) [6].

Se considera también que el medio interplanetario es transparente, y que los frentes de onda esféricos salidos de la fotosfera caen con el cuadrado de la distancia. Entonces las irradiancias espectrales $E_{top}(\lambda)$ en el tope de la atmósfera del planeta son:

$$E_{top}(\lambda) = \left(\frac{R}{r}\right)^2 E_{star}(\lambda), \qquad (2.2)$$

donde $R$ es el radio de Próxima del Centauro, estimada en unos 0.145 ± 0.011 $R_S$ [6] ($R_S$ es el radio del Sol) y $r$ es la distancia entre la estrella y el planeta, la cual se tomó como 0.1 unidades astronómicas (u.a.).

*2.2. Modelo atmosférico*   Se asumió una atmósfera similar a la de la Tierra actual, con un albedo $A_{in} = 0,4$ para la radiación infrarroja (que es la dominante en el espectro de Próxima del Centauro). Para hallar las irradiancias espectrales $E(\lambda,0^+)$ que llegan a la superficie del océano se utiliza la fórmula:

$$E(\lambda,0^+) = (1 - A_{in})E_{top}(\lambda). \qquad (2.3)$$

Parte de la radiación que llega a la superficie oceánica es reflejada, por lo que las irradiancias $E(\lambda,0^-)$ que llegan a la coordenada inmediatamente debajo de la superficie del océano

se calculan mediante:

$$E(\lambda,0^-) = [1-R_f]E(\lambda,0^+), \quad (2.4)$$

donde $R_f$ es el coeficiente de reflexión, obtenido de las fórmulas de Fresnel aplicado a la interferencia aire-agua. En este estudio se asume que el ángulo solar cenital es de cero grado, y se tiene entonces que $R_f \approx 0,02$.

Es conveniente además, multiplicar las irradiancias espectrales ultravioletas (UV) por el coeficiente de acción biológica para la inhibición de la fotosíntesis $\varepsilon(\lambda)$ [7] de la radiación ultravioleta, quedando de la siguiente forma:

$$E^*_{UV}(\lambda,0^-) = \varepsilon(\lambda)E(\lambda,0^-). \quad (2.5)$$

La función biológica $\varepsilon(\lambda)$ es un espectro de acción generalizado para la inhibición de la fotosíntesis y el daño del DNA del fitoplancton. El asterisco en las irradiancias ultravioletas significa que están ponderadas con un espectro de acción biológica.

*2.3. Modelo óptico oceánico* Las irradiancias para la luz fotosintéticamente activa que hay a diferentes profundidades $z$ en el océano se calculan según la ley de Lambert-Beer [8]:

$$E(\lambda,z) = E(\lambda,0^-) \cdot e^{-K(\lambda) \cdot z}, \quad (2.6)$$

donde $K(\lambda)$ es el coeficiente de atenuación de la luz para la longitud de onda $\lambda$.

Se usó la clasificación del agua oceánica de acuerdo a sus propiedades ópticas dadas por el oceanólogo N. Jerlov [4,9]. Las aguas claras son del tipo I, las intermedias son del tipo II y las aguas turbias son del tipo III [4,9]. En la referencia [4] se muestra un gráfico con los coeficientes de atenuación *vs.* longitud de onda para los tipos ópticos de aguas oceánicas.

*2.4. Patrón de circulación oceánica* Primero se calcularon las tasas de fotosíntesis para el océano en calma, y después se promediaron en celdas de 40 metros de profundidad (divididas en varias capas), considerando un patrón de corrientes verticales circulares (circulación de Langmuir).

*2.5. Modelo de Fotosíntesis* Para hallar las tasas de fotosíntesis se utilizó el modelo E para la fotosíntesis [10,11]:

$$\frac{P(z)}{P_S} = \frac{1-e^{-E_{LFA}(z)/E_S}}{1+E^*_{UV}(z)}, \quad (2.7)$$

donde $P(z)$ es la tasa de fotosíntesis del fitoplancton a la profundidad $z$, $P_S$ es la tasa de fotosíntesis máxima y:

$$E_{LFA}(z) = \sum_{\lambda_i}^{\lambda_f} E_i(\lambda,z)\Delta\lambda \quad (2.8)$$

y

$$E^*_{UV}(z) = \sum_{\lambda_i}^{\lambda_f} E^*_i(\lambda,z)\Delta\lambda \quad (2.9)$$

son las irradiancias totales para la luz fotosintéticamente activa (LFA, 400-1100 nm) y la radiación ultravioleta (UV, 280-399 nm) a la profundidad $z$. El parámetro $E_S$ mide la eficiencia de las especies en usar LFA: mientras más pequeño es, mayor es la eficiencia de las especies en usar LFA. En ausencia de luz ultravioleta, el denominador de la ecuación (2.7) toma el valor 1, de donde se desprende que $E_S$ representa la irradiancia de LFA para la cual la tasa de fotosíntesis alcanza el 63% de su valor máximo. Las especies de fitoplancton tienen un valor de $E_S$ en el rango (2 – 100) W/m². Para caracterizar este rango, en nuestros cálculos hemos tomado el valor máximo ($E_S = 100$ W/m²) y el mínimo ($E_S = 2$ W/m²) de este parámetro, que representa las especies menos eficientes y más eficientes, respectivamente, en usar LFA. En ambos casos $\Delta\lambda = 1$ nm. Es de destacar que se asume que los organismos en el exoplaneta podrían utilizar luz infrarroja para la fotosíntesis (hasta $\lambda_{máx} \approx 1100$ nm), no solo porque será la radiación electromagnética más común allí, sino porque en la Tierra ya se han descubierto unas pocas especies que fotosintetizan con el IR en el mencionado rango [12]. Además, longitudes de onda superiores podrían imponer requerimientos bioenergéticos mayores para lograr una cadena de transformación energía solar-energía química eficiente.

Se calculó después el promedio de la tasa de fotosíntesis $\left\langle \frac{P}{P_S} \right\rangle$ para las N capas que conforman la celda superior del océano, de 40 metros de profundidad (llamada capa límite, debido a que solo en ella se asume circulación). La tasa de fotosíntesis promedio resulta de:

$$\left\langle \frac{P}{P_S} \right\rangle = \frac{\sum_{n=1}^{N} \frac{P}{P_S}(n)}{N}. \quad (2.8)$$

### III. RESULTADOS Y DISCUSIÓN

En las Figuras 1-3 se muestran las tasas de fotosíntesis de los tres tipos de aguas I, II y III. Como podemos apreciar en los tres gráficos para los tres tipos de agua se alcanza el máximo de la tasa de fotosíntesis en la superficie del océano.

Estos gráficos son diferentes a los que se obtienen para cuando el planeta estudiado es la Tierra [13-15], básicamente debido a las diferencias en los espectros de emisión de ambas estrellas. El Sol emite bastante radiación en las bandas ultravioleta y visible. La radiación ultravioleta es una de las causantes de inhibir la fotosíntesis, ya que es una radiación más energética que el visible y el infrarrojo y tiende a dañar el ADN y el aparato fotosintético de las moléculas. Por eso en la superficie de los océanos de nuestro planeta no se alcanza un máximo de fotosíntesis, porque el fitoplancton esta más expuesto a estas radiaciones dañinas. El agua es un bloqueador de las radiaciones, entonces en capas de aguas por debajo de la superficie, donde ya han sido absorbidas las radiaciones más energéticas, el fitoplancton puede realizar mayor tasa de fotosíntesis.

A diferencia del Sol, Próxima del Centauro emite más radiación infrarroja que ultravioleta y visible, por eso la tasa de fotosíntesis alcanza el máximo en la superficie del océano, ya que prácticamente no hay radicación inhibitoria que dañe las biomoléculas.

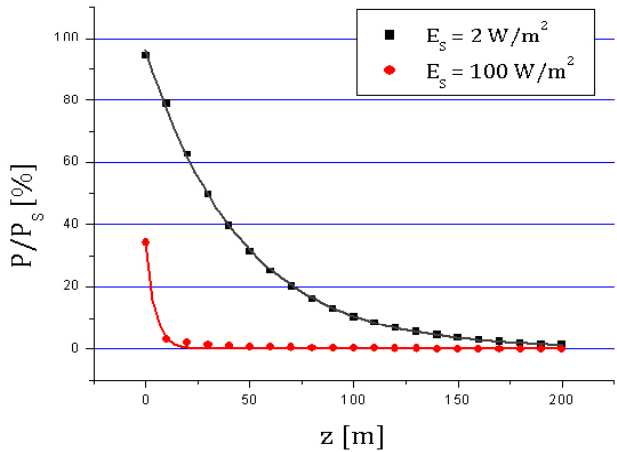

Fig. 1 Tasas de fotosíntesis en aguas tipo I (claras) *vs.* profundidad

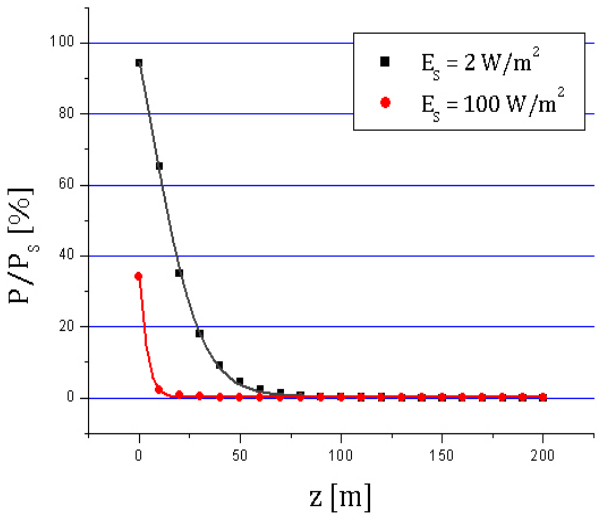

Fig. 2 Tasas de fotosíntesis en aguas tipo II (intermedias) *vs.* profundidad

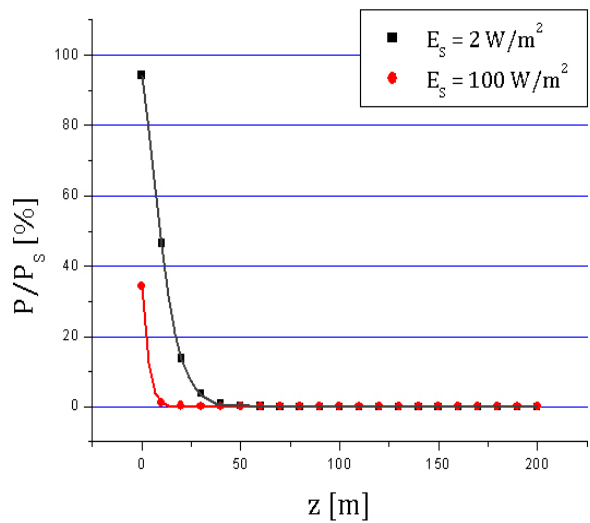

Fig. 3 Tasas de fotosíntesis en aguas tipo III (turbias) *vs.* profundidad

Los organismos muy poco eficientes en usar LFA ($E_S = 100$ W/m$^2$) presentan muy bajas tasas de fotosíntesis en los tres tipos de agua, mientras que los organismos más eficientes logran una tasa apreciable de fotosíntesis incluso a varias decenas de metros de profundidad.

En la tabla 1 se muestra el promedio de la tasa de fotosíntesis. Se obtiene mayor tasa de fotosíntesis en las aguas claras.

| Tabla 1 Promedio de la tasa de fotosíntesis. $E_S = 2$ W/m$^2$ | |
|---|---|
| Tipo de Aguas | $\left\langle \dfrac{P}{P_S} \right\rangle$ |
| Aguas Claras (tipo I) | 51,7 |
| Aguas Intermedias (tipo II) | 34,0 |
| Aguas Turbias (tipo III) | 22,7 |

Comparando con datos del planeta Tierra (Tabla 2) se observa que el promedio es considerablemente más alto que en el exoplaneta.

| Tabla 2 Promedio de la tasa de fotosíntesis para la Tierra. $E_S = 2$ W/m$^2$ | |
|---|---|
| Tipo de Aguas | $\left\langle \dfrac{P}{P_S} \right\rangle$ |
| Aguas Claras (tipo I) | 75,9 |
| Aguas Turbias (tipo III) | 79,1 |

Se obtiene para el planeta Tierra que el promedio de la tasa de fotosíntesis es más alto en aguas turbias que en aguas claras, esto se debe a que estos cálculos fueron hechos para los 40 primeros metros de profundidad del océano, y el UV inhibe la fotosíntesis en las primeras decenas de metros de profundidad para las aguas claras en la Tierra.

### IV. CONCLUSIONES

Los resultados sugieren que en principio podría ser posible la vida fotosintética en un exoplaneta que orbite una enana roja si los organismos allí presentes evolucionan la capacidad de utilizar el infrarrojo para fotosintetizar. La productividad obtenida es considerablemente menor que en la Tierra actual, pero no es baja. Para estimar la habitabilidad, hemos mostrado una arista esencialmente fotobiológica. Otras variables ambientales podrían jugar un rol importante, lo cual es objeto de futuros estudios en nuestro grupo.